\documentclass[twocolumn,a4paper]{article}

\usepackage[left=1.1cm,right=1.1cm,top=1.1cm,bottom=2cm]{geometry} 
\usepackage{graphicx} 
\usepackage{subcaption}
\usepackage{amsmath,amssymb}
\usepackage{mathtools} 
\usepackage{bm}
\usepackage[version=4]{mhchem}
\usepackage{braket}
\usepackage{booktabs}
\usepackage[separate-uncertainty = true,multi-part-units=single,bracket-ambiguous-numbers = false]{siunitx} 
\DeclareSIUnit\bar{bar}
\usepackage{microtype}
\usepackage[percent]{overpic}
\usepackage{xr-hyper}
\usepackage{changes}
\usepackage{comment}
\usepackage{subfiles}
\usepackage{titling}
\usepackage{float}
\usepackage{rccol}
\usepackage{lineno}

\usepackage[
  backend=biber,
  style=nature,
  url=false,
  doi=true,
  isbn=false
]{biblatex}

\usepackage{authblk}

\usepackage[hidelinks]{hyperref}
\usepackage{etoc}

\makeatletter
\renewcommand*{\url}[1]{\href{#1}{}}

\makeatother

\author[1]{Lukas Schleicher}
\author[2]{Leonard Geilen}
\author[2]{Alexander Musta}
\author[2]{Niklas Walter}
\author[1,3]{Anne Rodriguez}
\author[1]{Philipp Bredol}
\author[2,3]{Alexander W. Holleitner}
\author[1,3,4]{Eva M. Weig\thanks{eva.weig@tum.de}}

\affil[1]{Department of Electrical Engineering, School of Computation, Information and Technology, Technical University of Munich, 85748 Garching, Germany.}
\affil[2]{Walter Schottky Institute and Physics Department, Technical University of Munich, 85748 Garching, Germany.}
\affil[3]{Munich Center for Quantum Science and Technology (MCQST), Schellingstr.~4, 80799 Munich, Germany.}
\affil[4]{TUM Center for Quantum Engineering (ZQE), 85748 Garching, Germany.}

\title{Role of dissipation dilution in achieving high mechanical quality factors in van der Waals resonators}

\date{2026}

\begin{document}

\AddToHook{shipout/before}{\ifnum\value{page}=1 \global\hoffset=0.5cm \else\global\hoffset=0pt \fi}

\begin{refsection}

\maketitle
\begin{abstract}

Dissipation dilution has driven a revolution in the quality factors ($Q$) of top-down nanomechanical resonators. Yet it has not been leveraged for van der Waals (vdW) materials. Here we report quality factors up to $372{,}000\pm37{,}000$ in monolayer \ce{MoS2} drums at \SI{4}{\kelvin}, nearly an order of magnitude beyond the previous best for any vdW material at this temperature, enabled by a clean, polymer-free exfoliation technique. 
We develop a dissipation-dilution model for circular drums and analyze the contributing intrinsic and extrinsic dissipation mechanisms.
The \ce{MoS2} monolayer exhibits a dilution factor of order 
$1,000$, establishing sub-nanometer-thin resonators as ultimate dilution materials. The intrinsic quality factor, however, is limited by surface losses at atomic thickness.
Conversely, multilayer drums show a lower dilution factor but a high intrinsic quality factor, on par with established top-down materials of similar dimensions, both resulting from the increased thickness.
A two-level-system analysis of the temperature-dependent damping confirms this bulk-versus-surface picture and provides first insight into its microscopic origin. 
Our results identify dissipation dilution as the governing mechanism for high-$Q$ mechanical vdW resonators. Combined with strategies such as soft clamping or stress engineering, they establish atomically thin resonators as a platform for ultrahigh-$Q$ quantum transduction and sensing.

\end{abstract}

\subsection*{Introduction}
Over the past decade, dissipation dilution has emerged as a transformative paradigm in top-down nanomechanical engineering, enabling quality factors ($Q$) that would be unattainable through material optimization alone~\cite{sementilliNanomechanicalDissipationStrain2022, engelsenUltrahighqualityfactorMicroNanomechanical2024}. In strained mechanical resonators, the concurrence of tensile stress and geometric nonlinearity dramatically increases the quality factor: while the static tensile stress increases the stored elastic energy of the vibrating resonator, the dissipated energy remains largely unaffected. This effectively dilutes the intrinsic loss. The concept was first exploited in the mirror suspensions of gravitational-wave interferometers~\cite{saulsonThermalNoiseMechanical1990,gonzalezBrownianMotionMass1994}, where pendulum modes benefit from an analogous dilution of internal friction by the conservative gravitational restoring force. Translated to the nanoscale, dissipation dilution was then recognized as the origin of the anomalously high quality factors observed in tensioned silicon nitride nanostructures~\cite{verbridgeHighQualityFactor2006,unterreithmeierDampingNanomechanicalResonators2010, yuControlMaterialDamping2012,villanuevaEvidenceSurfaceLoss2014}, establishing a first generation of high-$Q$ nanomechanical resonators. Building on this foundation, soft clamping~\cite{tsaturyanUltracoherentNanomechanicalResonators2017} and strain engineering~\cite{ghadimiElasticStrainEngineering2018} were subsequently developed as strategies to push dilution further: soft clamping engineers the mode profile so that bending, and thus flexural dissipation, is suppressed at the vibration antinodes near the clamps, while strain engineering tailors the local stress field to maximize the tensile energy throughout the resonator body.
Together, these advances have raised room-temperature quality factors by several orders of magnitude~\cite{sementilliNanomechanicalDissipationStrain2022, engelsenUltrahighqualityfactorMicroNanomechanical2024}, with the prospect of quantum coherent operation of a nanomechanical resonator at room temperature now within reach~\cite{tsaturyanUltracoherentNanomechanicalResonators2017, ghadimiElasticStrainEngineering2018}.

Dissipation dilution in thin-film materials mostly relies on stoichiometric and thus high-stress amorphous silicon nitride (\ce{Si3N4})~\cite{verbridgeHighQualityFactor2006,unterreithmeierDampingNanomechanicalResonators2010,yuControlMaterialDamping2012,villanuevaEvidenceSurfaceLoss2014,tsaturyanUltracoherentNanomechanicalResonators2017,ghadimiElasticStrainEngineering2018}, though the paradigm has more recently been extended to amorphous silicon carbide (\ce{a-SiC})~\cite{xuHighStrengthAmorphousSilicon2024} and to crystalline materials including \ce{SiC}~\cite{romeroEngineeringDissipationCrystalline2020}, \ce{Si}~\cite{beccariStrainedCrystallineNanomechanical2022}, \ce{InGaP}~\cite{manjeshwarHighQTrampolineResonators2023}, and \ce{AlN}~\cite{ciersNanomechanicalCrystallineAlN2024}. 
Two-dimensional (2D) materials, by contrast, have yet to benefit from this revolution. While their atomically thin membranes offer extraordinary mechanical properties, including high intrinsic strength, low mass, and gate-tunable tension, a comprehensive theoretical framework connecting dissipation mechanisms to material and device parameters, and in particular to the role of dissipation dilution, remains absent.  
This connection gap is particularly striking given a fundamental scaling argument: dissipation dilution grows with decreasing resonator thickness, and a monolayer — at the physical minimum of one molecular plane — represents the ultimate limit of this scaling. A monolayer can therefore be regarded as the \textit{ultimate dilution material}, a platform in which the geometric suppression of bending rigidity relative to tensile energy is maximized by definition. Realizing dissipation-diluted quality factors in such a system would open transformative possibilities for vdW-material-based sensing~\cite{lemmeNanoelectromechanicalSensorsBased2020,  aguilaPhotothermalResponsivityVan2022} and quantum transduction with high efficiency~\cite{gaoHighspeedQuantumTransducer2020}.

\begin{figure}[t]
    \centering 
    \begin{overpic}[width=8.99cm]{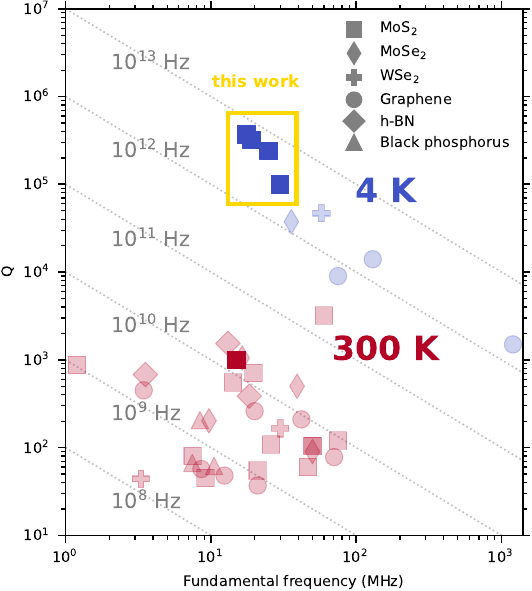}
    \end{overpic}
    \caption{\textbf{Overview of van der Waals mechanical resonators in the literature.} 
    Quality factor vs. fundamental eigenfrequency for mechanical resonators made of different van der Waals materials, detailed in the legend. Data points measured at room temperature (\SI{4}{\kelvin}) are indicated in red (blue). Data points obtained in this work are highlighted in yellow. Gray lines indicating the $Q \cdot f$ product serve as a guide to the eye.
    See Supplementary Fig.~\ref{fig:LitOverview} for details of all references.}
    \label{fig:fig0}
\end{figure}

Here we report — to the best of our knowledge — the highest mechanical quality factors measured to date in any van der Waals material resonator, reaching values of $Q = 372{,}000 \pm 37{,}000$ for the fundamental mode of a monolayer \ce{MoS2} drum at \SI{4}{K}. Figure~\ref{fig:fig0} puts these results in perspective with previous work between \SI{4}{K}~\cite{chenPerformanceMonolayerGraphene2009,morellOptomechanicalMeasurementThermal2019,zandeLargeScaleArraysSingleLayer2010,jungGHzNanomechanicalResonator2019} and room temperature~\cite{bunchElectromechanicalResonatorsGraphene2007a,leeHighFrequencyMoS22013a,pittsEvidencingDissipationDilution2024,kramerStraindependentDampingNanomechanical2015,wopereisTuningDissipationDilution2024}.
We show that our record values are quantitatively explained by a rigorous theoretical model for dissipation dilution in circular membrane resonators, applied to the ultimate limit of a monolayer material, combined with exceptionally low intrinsic material damping. The latter is attributed to a recently developed polymer-free,
gold-assisted exfoliation technique~\cite{velickyMechanismGoldAssistedExfoliation2018, liuDisassembling2DVan2020, geilenSituMappingPressureInduced2025}, giving rise to ultralow damping rates when cooled to temperatures below \SI{50}{K}. 

The monolayer results are complemented by the investigation of multilayer drums consisting of a few tens of layers. As expected, we find a lower dilution factor due to the increased thickness. However, the multilayer drums feature significantly higher intrinsic quality factors of $Q_\mathrm{int} \approx 10{,}000$, almost $50$ times that of the monolayer, and notably on par with typical top-down dissipation-dilution materials~\cite{hochreiterMonolithic4HSiCNanomechanical2025}. This is attributed to a remarkably low material dissipation in the volume of multilayer \ce{MoS2} drums, mitigating the higher losses arising on the drum surface, which limit the performance of the monolayers. The high intrinsic quality factor of the multilayer drum is thus a consequence of its increased thickness, which simultaneously reduces the dilution factor. The trade-off between the diluted and intrinsic quality factors in multilayers highlights the established role of the surface-to-volume ratio on the intrinsic damping~\cite{villanuevaEvidenceSurfaceLoss2014}.  
The comparison of mono- and multilayer drums showcases the interplay of dissipation dilution and the intrinsic, undiluted quality factor in maximizing the measured quality factor.
Together, our results establish dissipation dilution as the governing mechanism for high-$Q$ mechanical response in van der Waals resonators and provide a quantitative framework for engineering quality factors in this class of materials.

\subsection*{Dissipation dilution in circular van der Waals drums}
The quality factor of a mechanical resonator is defined as $2\pi$ times the ratio of the total stored energy to the total energy lost per oscillation cycle. 
In general, it combines both intrinsic and extrinsic loss mechanisms. We first isolate the intrinsic part, which is related to structural dissipation within the resonator and is the dominant contribution in this work.
For a tensioned membrane, this results in a diluted intrinsic quality factor given by
\begin{equation}
    Q_{\mathrm{dil}} = 2\pi
    \frac{W_{\mathrm{ten}} + W_{\mathrm{flex}} + W_{\mathrm{elong}}}
    {\Delta W_{\mathrm{flex}} + \Delta W_{\mathrm{elong}}} ,
    \label{eq:energyratio} 
    \end{equation}
where $W_{\mathrm{ten}}$, $W_{\mathrm{flex}}$, and $W_{\mathrm{elong}}$ are the tensile, flexural, and elongational contributions to the stored elastic energy, and $\Delta W_{\mathrm{flex}}$ and $\Delta W_{\mathrm{elong}}$ are the corresponding dissipated energies per cycle. The tension arises from a conservative potential, and therefore contributes to the stored energy without introducing additional dissipation~\cite{fedorovGeneralizedDissipationDilution2019a}, -- the origin of dissipation dilution.
Equation~(\ref{eq:energyratio}) separates into a dilution factor
$D$ and an intrinsic quality factor $Q_\mathrm{int}$, accounting for the intrinsic losses in the (unstressed) material~\cite{sementilliNanomechanicalDissipationStrain2022}
\begin{equation}
     Q_\mathrm{dil} = D \cdot \, Q_\mathrm{int} \quad \mathrm{with}
     \label{eq:dilution}
\end{equation}%
\begin{equation}
    \label{eq:DilutionFactor}
    D = \frac{W_{\mathrm{ten}} + W_{\mathrm{flex}} +
    W_{\mathrm{elong}}}{W_{\mathrm{flex}} + W_{\mathrm{elong}}} \approx 1+\frac{W_{\mathrm{ten}}}{W_{\mathrm{flex}}} .
\end{equation}

The contribution $W_{\mathrm{elong}}$ from the bending-induced elongation is negligible in the linear response regime (see Supplementary Sections~\ref{SI:Amplitude} and~\ref{fig:energys}). The dilution factor, therefore, quantifies how much the tensile energy stored in the resonator ``dilutes'' the contribution of flexural losses in the quality factor.
Rigorous theoretical models of \textit{D} have so far been developed exclusively for quasi one-dimensional strings and rectangular membranes, which are the natural geometries of top-down fabricated nanostructures~\cite{saulsonThermalNoiseMechanical1990,gonzalezBrownianMotionMass1994,unterreithmeierDampingNanomechanicalResonators2010,schmidDampingMechanismsHigh$Q$2011,yuControlMaterialDamping2012,fedorovGeneralizedDissipationDilution2019a}. The geometry of a fully clamped circular drum — standard for van der Waals resonators — requires a separate treatment, as the axisymmetric mode shapes and their associated stress and strain fields take a fundamentally different form.

In polar coordinates, the stored tensile and flexural energy are defined as
\begin{equation}
    W_{\mathrm{ten}}
    = \sigma h \pi \int\limits_0^{R}
    \left(\frac{\partial u}{\partial r}\right)^2 r\,\mathrm{d}r,
\end{equation}
\begin{equation}
    W_{\mathrm{flex}}
    = \pi \kappa \int\limits_0^{R}
    \left[\left(\frac{\partial^2 u}{\partial r^2} + \frac{1}{r}\frac{\partial u}{\partial r}\right)^2
    - 2(1-\nu)
    \frac{\partial^2 u}{\partial r^2}
    \frac{1}{r}\frac{\partial u}{\partial r}\right] r\,\mathrm{d}r,
\end{equation}
respectively. The radius and thickness of the drum are denoted by $R$ and $h$. We use $u(r)$ as the radial component of the mode shape, and $\sigma$, $E$, $\nu$, and $\kappa = E \cdot h^3/\left[12 (1-\nu^2)\right]$ represent the tensile stress, Young's modulus, Poisson ratio, and flexural rigidity, respectively. All material parameters used in this work are listed in Supplementary Table~\ref{tab:material parameters}. 

In van der Waals material-based mechanical resonators, correlations between tension and an increasing quality factor have been observed, but mostly been attributed to the reduction of the dissipative effects of wrinkles~\cite{kramerStraindependentDampingNanomechanical2015, xieStrainingTuningAtomic2021a} 
or to thermoelastic effects~\cite{zhangStrainModulatedDissipationTwoDimensional2022,zhangStrainDilutionThermoelastic2026,zhangThermallymodulatedDissipationDilution2026}.
Only a few studies have been interpreted as signatures of dissipation dilution~\cite{wopereisTuningDissipationDilution2024,pittsEvidencingDissipationDilution2024}, using an approximation for tensioned plates in the membrane-plate crossover regime to express the dilution factor as being proportional to the square of the eigenfrequency, $Q \propto f_0^2$~\cite{steenekenDynamics2DMaterial2021}.
To date, a rigorous and quantitative model applicable to the ultralow-bending-rigidity regime of monolayer and few-layer drums has remained elusive. Here, we present a model developed from elastic theory  that evaluates the dilution factor using the full energy balance in Eq.~(\ref{eq:DilutionFactor}) for the exact axisymmetric mode profiles of the circular drum as a function of tension, thickness, and radius. The mode shape integrals entering this calculation do not yield a compact closed form, and the result is obtained numerically; a detailed derivation is provided in Supplementary Section~\ref{SI:DilutionCalculus}. We find that the dilution factor is highly sensitive to resonator geometry, tension, and mode shape, and that it increases with decreasing thickness, making atomically thin membranes particularly favorable hosts for dissipation dilution.

\subsection*{Mechanical characterization of \ce{MoS2} drums}
\begin{figure}[t]
    \centering 
    \begin{overpic}[width=9cm]{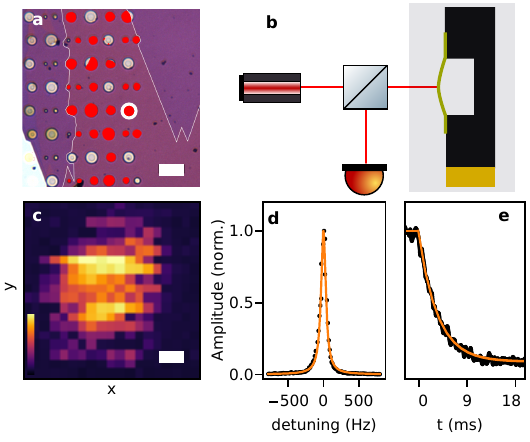}
    \end{overpic}
    \caption{\textbf{Experimental setup.} \textbf{(a)} Optical microscope image of \ce{Si/SiO_2} substrate coated with Ti/Au and transferred \ce{MoS2}, overlayed with a photoluminescence map indicating the position of suspended, luminescent monolayers (red). Scalebar corresponds to $25\,\text{µm}$. \textbf{(b)} Interferometric measurement scheme. The suspended drum is driven with a piezo actuator (yellow box) and exposed to a focused infrared laser spot. The drum's vibration modulates the reflected light, which is detected by a photodetector. \textbf{(c)-(e)} Fundamental mechanical mode of a monolayer drum with a radius of \SI{5.5}{\micro\meter} (marked by a white circle in (a)) at $\SI{4}{K}$. \textbf{(c)} Mode shape, depicting the maximum amplitude as a function of position. Scale bar corresponds to \SI{2}{\micro\meter}; color scale indicates normalized amplitude. \textbf{(d)} Linear response of the fundamental mode at \SI{18.1}{MHz} taken in the center of the drum along with fit (orange) and \textbf{(e)} ringdown measurement taken at the same position including exponential fit (orange), yielding a $Q = 229{,}000 \pm 14{,}000$.}
    \label{fig:fig1}
\end{figure}

We study suspended \ce{MoS2} drums hosted on several chips, comprising both monolayer (1L) and multilayer (ML) membranes with radii ranging from \SI{2.5}{\micro\meter} to \SI{6}{\micro\meter}. Figure~\ref{fig:fig1}a shows a microscope image of one of the measured chips, with the photoluminescence overlay identifying the monolayer drums~\cite{makAtomicallyThinMoS2010}. For the monolayer, we assume a thickness of \SI{0.65}{\nano\meter} (see Supplementary Table~\ref{tab:material parameters}). The layer thickness of the multilayers is independently verified by atomic force microscopy. For the multilayer drums discussed below, it is \SI{24}{\nano\meter}, corresponding to approximately $37$ layers. Mechanical characterization is performed using piezo-actuated optical interferometry, as sketched in Fig.~\ref{fig:fig1}b. All measurements are carried out in a variable-temperature cryostat between room temperature and \SI{4}{\kelvin}, at chamber pressures below \SI{e-4}{\milli\bar} (\SI{e-8}{\milli\bar}) at room (cryogenic) temperature.

Figure~\ref{fig:fig1}c-e shows the linear mechanical response of the fundamental mode of a \SI{5.5}{\micro\meter} radius monolayer drum (marked by a white circle in Fig.~\ref{fig:fig1}a) measured at \SI{4}{\kelvin}. Figure~\ref{fig:fig1}c displays the spatial mode profile. It is obtained by scanning the laser spot across the drum area, taking a response spectrum at each position. We ensure that each spectrum is measured in the linear-response regime and fit with a Lorentzian to extract the resonance frequency. The data in Fig.~\ref{fig:fig1}c depicts the peak amplitude of each fit. The resulting circular pattern, with an amplitude maximum at the center, clearly identifies the fundamental mode. The complete frequency response measured in the center of the drum is shown in Fig.~\ref{fig:fig1}d, along with a Lorentzian fit yielding an eigenfrequency of \SI{18.1}{MHz} at \SI{4}{\kelvin}. The fundamental resonance frequencies of all investigated devices lie between \SI{16}{\mega\hertz} and \SI{35}{\mega\hertz}, depending on their geometry. With higher-order modes resolved up to \SI{45}{\mega\hertz}; we attribute the upper frequency limit to the low-pass roll-off of the piezo shaker's capacitive load.
A ringdown measurement taken at the same point is depicted in Fig.~\ref{fig:fig1}e with an exponential fit. Quality factors are extracted from both the linewidth of the driven-frequency response and ringdown measurements. In general, ringdown measurements provide a more reliable method to determine $Q$, since they are immune to the spectral diffusion caused by frequency fluctuations, which can artificially broaden the apparent linewidth and thus underestimate $Q$ in swept-frequency spectra~\cite{schneiderObservationDecoherenceCarbon2014}. 
In case ringdown measurements are not feasible, spectral measurements serve as a lower bound of the quality factor (see Supplementary Section~\ref{SI:Error}).

Across all investigated drum geometries, quality factors at base temperature range from $40{,}000$ to $372{,}000$ (c.f. Supplementary Fig.~\ref{fig:Measurement}b,c). For the drum shown in Fig.~\ref{fig:fig1}c,d, we obtain $Q = 229{,}000 \pm 14{,}000$ at \SI{4}{\kelvin}. 
These values represent the highest quality factors reported to date in any 2D material resonator — and, for \ce{MoS2} specifically, exceed the previous best by nearly an order of magnitude (see Fig.~\ref{fig:fig0} and Supplementary Section~\ref{SI:LiteratureComparison}).

The quality factor is decomposed into its diluted intrinsic and its extrinsic contribution
\begin{equation}
    \frac{1}{Q} =
    \frac{1}{Q_\mathrm{dil}} + \frac{1}{Q_\mathrm{ext}} = 
    \frac{1}{D}\cdot \frac{1}{Q_\mathrm{int}} + \frac{1}{Q_\mathrm{ext}}, 
    \label{eq:Qdecomposition}
\end{equation}
where $Q_\mathrm{int}^{-1}=\sum_i Q_{\mathrm{int},i}^{-1}$ and 
$Q_\mathrm{ext}^{-1}=\sum_i Q_{\mathrm{ext},i}^{-1}$ sum over all intrinsic 
and extrinsic loss mechanisms, respectively. Only the intrinsic terms are 
subject to dissipation dilution, as stated in Eq.~(\ref{eq:dilution}).
Among the extrinsic contributions, clamping loss is the most relevant to 
our analysis, as it depends sensitively on drum geometry, substrate, and 
tension~\cite{wilson-raeHighNanomechanicsDestructive2011}, see Supplementary Section~\ref{SI:Damping}. 
We find that clamping loss is negligible for the monolayers, but needs to be accounted for the multilayer drums. Gas damping~\cite{leeAirDampingAtomically2014} is negligible at the base pressure of the vacuum chamber ($<\SI{e-4}{\pascal}$) and becomes even more so at cryogenic temperatures due to cryopumping~\cite{pobellMatterMethodsLow2007,beccariStrainedCrystallineNanomechanical2022}. Electrostatic transducer losses~\cite{matisEnergyDissipationPathways2017} are entirely absent as the investigated drums are not gated and no voltage is applied across the chip. 

All in all, for monolayer drums, the measured $Q$ is therefore governed entirely by intrinsic mechanisms, and coincides with the diluted intrinsic $Q_\mathrm{dil} = D \cdot Q_\mathrm{int}$. Equivalently, the measured damping rate $\Gamma =2\pi f_0 / Q$ 
reflects purely intrinsic losses. For multilayer drums, the clamping loss contribution has to be subtracted from the measured quality factor in order to analyze $Q_\mathrm{dil}$.

\subsection*{Dissipation dilution at \SI{4}{\kelvin}}
\begin{figure}[t]
    \centering
    \begin{overpic}[width=9cm]{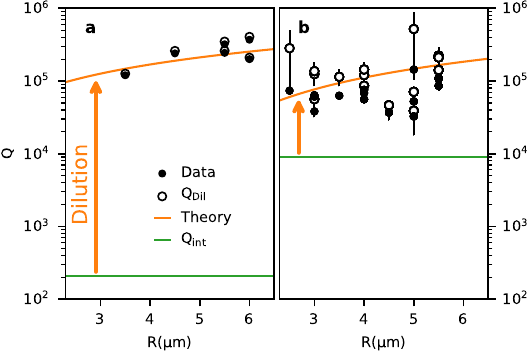}
    \end{overpic}
    \caption{\textbf{Dissipation dilution and drum geometry}. Quality factors obtained from ringdown measurements at \SI{4}{\kelvin} in \textbf{(a)} monolayer (\SI{0.65}{\nano\meter}, black dots) and \textbf{(b)} multilayer drums ($37$ layers, \SI{24}{\nano\meter}, black dots) for different drum radii. Each data point is averaged over multiple ringdowns acquired on the same drum. The error bars of the monolayer drums are smaller than the symbol size. The diluted intrinsic $Q_{\mathrm{dil}} = D \cdot Q_{\mathrm{int}}$ (open circle
    ) are obtained by subtracting the extrinsic contribution of clamping loss, and error bars of propagated error are shown.
    Fits of the dissipation dilution model are shown as orange lines. The fits yield an intrinsic quality factor of $Q_{\mathrm{int}} = 212$ for the monolayer  and $Q_{\mathrm{int}} = 9270$ for the multilayer drums (green lines). Orange arrows visualize the dilution factor.}
    \label{fig:fig4}
\end{figure}

We now turn to the role of dissipation dilution in governing the quality factors of our \ce{MoS2} drums. Figure~\ref{fig:fig4} shows the measured $Q$ at base temperature of \SI{4}{\kelvin} as a function of drum radius for two different thicknesses. We compare monolayer drums of thickness \SI{0.65}{\nano\meter} (Fig.~\ref{fig:fig4}a) and multilayer drums with a thickness of \SI{24}{\nano\meter} (Fig.~\ref{fig:fig4}b). 
All quality factors are extracted from ringdown measurements. Each data point represents an average over the ringdown measurements performed on a given drum; multiple data points at the same radius correspond to independent drums of identical geometry. Error bars reflect the uncertainty of the ringdown fits; they are, in many cases, smaller than the symbol size (see Supplementary Section~\ref{SI:Error} for detailed discussion on error estimation). From the measured quality factor (circles), we compute the diluted intrinsic $Q_{\mathrm{dil}} = D \cdot Q_{\mathrm{int}}$ (open circles) by subtracting the contribution of extrinsic clamping loss obtained from Eq.~(\ref{eq:clamping}) in Supplementary Section~\ref{SI:clamping loss}. For the monolayer drums in Fig.~\ref{fig:fig4}a, the open circles and open circles almost coincide, indicating that extrinsic clamping loss is negligible. For the multilayer drums in Fig.~\ref{fig:fig4}b, the open circles lie above the circles, revealing the non-negligible influence of clamping loss. 

Next, we fit the dissipation dilution model to $Q_{\mathrm{dil}}$ for both mono- and multilayer drums (orange lines). The fits yield intrinsic quality factors of $Q_\mathrm{int} = 212 \pm 10$ for the monolayer and $Q_\mathrm{int} = 9,270\pm 400$ for the multilayer drums (green lines). 
We would like to draw the reader's attention to the pronounced difference between the mono- and the multilayer results. The $37$-layer drum exhibits an 
intrinsic quality factor of $Q_\mathrm{int} \approx 10{,}000$, 
on par with benchmark top-down dissipation-dilution materials at comparable thickness, such as stoichiometric \ce{Si3N4} ($Q_\mathrm{int} \approx 2,800$ at $h=$\SI{30}{\nano\meter}~\cite{villanuevaEvidenceSurfaceLoss2014}) or crystalline \ce{Si} ($Q_\mathrm{int} \approx 8,400$ at $h=$\SI{14}{\nano\meter}~\cite{beccariStrainedCrystallineNanomechanical2022}).
At \SI{24}{\nano\meter}, the multilayer is still ultrathin by conventional nanomechanical standards, yet thick enough for the volume contribution to dominate over surface losses, resulting in a low intrinsic damping. The dilution factor (visualized by an orange arrow), however, is moderate — between $10$ and $20$ depending on drum radius — reflecting the comparatively larger thickness. The monolayer presents a strikingly different picture: its sub-nanometer thickness of only \SI{0.65}{\nano\meter} renders it exquisitely sensitive to surface losses, resulting in a much lower intrinsic 
quality factor of $Q_\mathrm{int} \approx 200$. Yet this same thickness drives a dilution factor of approximately $1,000$ (orange arrow) — about $50$ times that of the 
multilayer, consistent with the scaling 
argument put forward in the introduction. It is this interplay between a reduced intrinsic $Q_\mathrm{int}$ and a greatly enhanced dilution factor that determines 
the overall quality factor, and underscores the role of the monolayer as the ultimate dilution material.

\subsection*{Temperature dependence of the quality factor}
\begin{figure}[t]
    \centering
\begin{overpic}[width=\columnwidth]{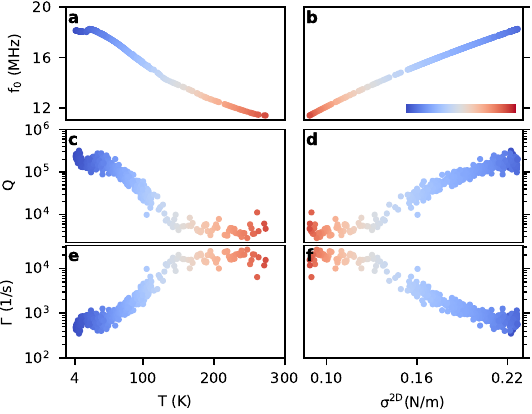}
    \end{overpic}
    \caption{\textbf{Temperature dependence of mechanical properties.} \textbf{(a,b)} Fundamental eigenfrequency of a monolayer drum with radius of \SI{5.5}{\micro\meter} as a function of temperature and tension, respectively. The tension is obtained from the measured eigenfrequency using Eq.~\ref{eq:membranefreq}.  \textbf{(c,d)} Quality factor extracted from the linear response of the drum as a function of temperature and tension, respectively. \textbf{(e,f)} Damping rate as a function of temperature and tension, respectively. The color of the datapoints represents temperature from room temperature (red) down to \SI{4}{\kelvin} (blue).}
    \label{fig:fig3}
\end{figure}

The full temperature dependence of the mechanical response of a monolayer drum with a radius of \SI{5.5}{\micro\meter} (c.f. Fig.~\ref{fig:fig1}c-e) from \SI{4}{\kelvin} to room temperature is shown in Fig.~\ref{fig:fig3}.
We begin with the eigenfrequency, depicted as a function of temperature in Fig.~\ref{fig:fig3}a. The eigenfrequency serves as a direct measure of the drum 
tension and thus provides the key input parameter for the dissipation dilution 
model. Monolayer \ce{MoS2} drums operate in the membrane regime, where the 
restoring force is dominated by the two-dimensional tension $\sigma^{\mathrm{2D}} = \sigma \cdot h$ rather than the bending rigidity $\kappa$, as the extraordinarily small thickness ensures that the dimensionless tension $\sigma^{\mathrm{2D}} R^2 / \kappa \gg 
1$ even at moderate tensions~\cite{castellanos-gomezSingleLayerMoS22013, 
leeHighFrequencyMoS22013a}. The fundamental mode frequency then takes the form
\begin{equation}
    f_0 \approx \frac{2.404}{2\pi R} \sqrt{\frac{\sigma^{\mathrm{2D}}}{\rho h}},
    \label{eq:membranefreq}
\end{equation}
where $2.404$ is the approximate first zero of the Bessel function $J_0$, $R$ is the drum 
radius, $\sigma$ the three-dimensional stress, and $\rho$ the mass density.
The fundamental frequency
decreases monotonically with increasing temperature. We attribute this to a thermally induced decrease in tensile stress, arising from the larger thermal expansion coefficient of 
\ce{MoS2} relative to the silicon substrate~\cite{zhangThermalExpansionCoefficient2019, 
cartamil-buenoMechanicalCharacterizationCleaning2017} (see Supplementary 
Section~\ref{SI:thermal Expansion} and Supplementary Fig.~\ref{fig:drifts}b). 
The tension is extracted from the measured eigenfrequencies of Fig.~\ref{fig:fig3}a using Eq.~(\ref{eq:membranefreq}). The measured eigenfrequencies are re-plotted as a function of tension in Fig.~\ref{fig:fig3}b, depicting the square-root dependence of Eq.~(\ref{eq:membranefreq}).

Figures~\ref{fig:fig3}c,d show the corresponding temperature dependence of the quality factor $Q$, again plotted both as a function of temperature and as a function of tension, respectively. The quality factor is obtained from the spectral linewidth via $Q = 2\pi f_0 / \Gamma$ across the temperature-dependent measurements in Fig.~\ref{fig:fig3}, as ringdown measurements become impractical at elevated temperatures due to insufficient signal-to-noise ratio. The validity of this approach is confirmed by the agreement between spectral and ringdown measurements at low temperatures, see Supplementary Section~\ref{SI:Error}.
Figure~\ref{fig:fig3}e,f depicts the spectral linewidth $\Gamma$ as a function of temperature and tension, respectively.

Figure \ref{fig:fig3} demonstrates that the temperature dependence of $Q$ cannot be explained by the evolution of the eigenfrequency or of the dissipation rate alone.
This suggests that not only dissipation dilution arising from 
the temperature-induced change in tension (Figs.~\ref{fig:fig3}a,b) but also a temperature-dependent 
damping rate (Figs.~\ref{fig:fig3}e,f) contribute to the observed temperature dependence of $Q$  (Figs.~\ref{fig:fig3}c,d). 
This motivates a closer examination of both effects, which we address in the following to disentangle their contribution to the measured $Q$. 

\begin{figure*}[t]
    \centering
    \begin{overpic}[width=18cm]{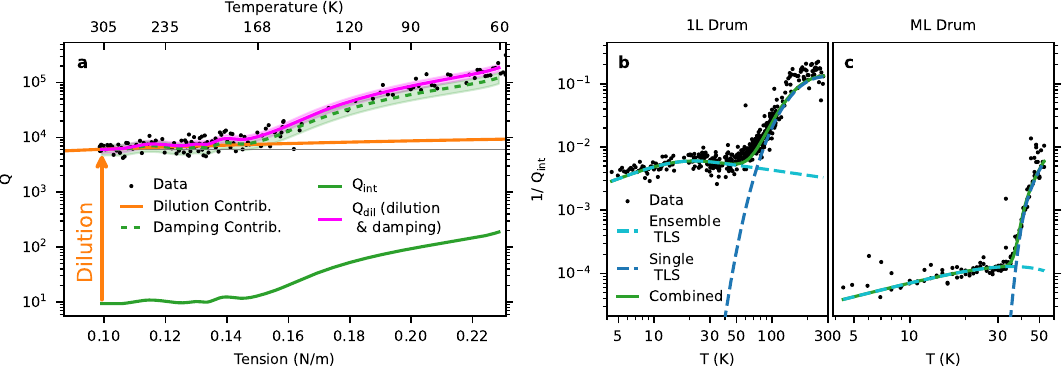}
    \end{overpic}
    \hfill
    \caption{\textbf{Influence of temperature on dissipation dilution and on damping.}  \textbf{(a)} Quality factor of a monolayer drum with \SI{5.5}{\micro\meter} radius as a function of temperature-induced tension (black dots). The orange solid and green dashed lines depict the contribution of the temperature-dependent dilution factor and damping rate to the quality factor, starting from the room-temperature quality factor as the baseline. The orange arrow visualizes the dilution factor at room temperature. The green solid line shows the intrinsic quality factor. The pink line combines both contributions, yielding excellent agreement with the measured data. 
    Shaded regions indicate error corridors.
    \textbf{(b,c)} Inverse intrinsic quality factor of  \textbf{(b)} the same monolayer as in \textbf{(a)} and  \textbf{(c)} a $37$-layer drum with \SI{5}{\micro\meter} radius
    as a function of temperature. The green solid lines depict fits of the combined TLS model. The underlying single-defect and defect-ensemble contributions representing the bulk and surface-related dissipation are included as blue and turquoise dashed lines, respectively.
   }
    \label{fig:fig5}
\end{figure*}

Having established the interplay between the dilution factor $D$ and the intrinsic quality factor $Q_\mathrm{int}$ at base temperature,  we now examine the temperature dependence of $Q$ to elucidate the physical origin of both contributions (Fig.~\ref{fig:fig5}).
Figure~\ref{fig:fig5}a displays $Q$ as a function of tension for a monolayer drum with a radius of \SI{5.5}{\micro\meter} (same as in Figs. \ref{fig:fig1}c-e and ~\ref{fig:fig3}), with temperature correspondingly decreasing from left to right. 
As before, the measured eigenfrequency is used to convert the temperature axis into tension using Eq.~(\ref{eq:membranefreq}). 
Raw data of the measured $Q$, here assumed to be equal to $Q_\mathrm{dil}= D \cdot Q_\mathrm{int}$ (see Fig.~\ref{fig:fig4}a), are shown as black dots. We use the quality factor measured at room temperature, $Q(\mathrm{RT})=6,000$, as the baseline of the analysis.
We then isolate the contribution of the temperature-dependent damping rate $\Gamma(T)$, extracted from the spectral linewidth, to the evolution of $Q$ by assuming that the tensile stress, and thus the eigenfrequency, remains fixed at its room-temperature value. The resulting damping contribution of the quality factor is obtained by scaling as $Q(\mathrm{RT}) \cdot \Gamma(\mathrm{RT})/\Gamma(T)$ (green dashed line). 
This curve accounts for a change by more than one order of magnitude in $Q$ between base and room temperature, reflecting the strong temperature dependence of the intrinsic damping rate.

Next, we isolate the contribution of dissipation dilution to the data. Using the same room temperature baseline, we fit the dissipation dilution model to this datapoint, assuming the damping remains fixed at its room-temperature value $\Gamma(\mathrm{RT})$. This procedure allows us to establish the temperature dependence of the dilution factor $D(T)$, and thus the dilution contribution to the quality factor (orange solid line). 
The dilution factor at room temperature is close to $650$ (orange arrow). The dilution factor at base temperature increases by $\approx  50 \% $ as a result of further temperature-induced tensioning. In total, the dilution factor at base temperature amounts to approx. $1,000$,  underscoring the dominant role of dissipation dilution in achieving the observed record $Q$ values. 

Finally, the dilution model also reveals the intrinsic quality factor at room temperature to be $Q_\mathrm{int}(\mathrm{RT})=9$. 
This allows us to obtain the temperature dependence of the intrinsic quality factor by scaling as $Q_\mathrm{int}(T)=Q_\mathrm{int}(\mathrm{RT}) \cdot \Gamma(\mathrm{RT})/\Gamma(T)$ 
(green solid line, we use a Gaussian filter on the data for better illustration). With decreasing temperature, $Q_\mathrm{int}$ increases to $186 \pm 19$ for this particular drum. Combining this temperature-dependent intrinsic $Q_\mathrm{int}(T)$ with the obtained dilution factor $D(T)$ yields the total quality factor (pink line accompanied by a light-pink error corridor accounting for an error of $10\%$ of the Q factor). It is found to be in excellent agreement with the raw data across the full temperature range, demonstrating that the temperature dependence of $Q$ is quantitatively accounted for by two effects: a temperature-induced change in tension that modifies the dissipation dilution factor and a temperature-dependent intrinsic damping rate. While both contributions are significant, the overall $Q$ is dominated by dissipation dilution, which elevates $Q$ by nearly three orders of magnitude relative to the intrinsic value.

\subsection*{Intrinsic damping}
Having established the role of dissipation dilution, we now turn to the temperature dependence of the intrinsic quality factor and the underlying intrinsic damping processes~\cite{imbodenDissipationNanoelectromechanicalSystems2014}. Possible intrinsic damping mechanisms include loss arising from defects in the material, surface-related losses, thermoelastic damping, and phonon-phonon scattering losses in the Akhiezer regime (see Supplementary Section~\ref{SI:Damping} for a more detailed discussion of all dissipation mechanisms). We estimate thermoelastic damping~\cite{lifshitzThermoelasticDampingMicro2000,metcalfThermoelasticDampingMicromechanical2009} to lie more than $9$ orders of magnitude (one order of magnitude) below the observed intrinsic damping for the monolayer (multilayer). Phonon–phonon scattering~\cite{marisInteractionSoundWaves1971} — evaluated in the Akhiezer regime, since $\omega\tau_\mathrm{ph} \ll 1$ for the thermal-phonon relaxation time at all measured temperatures — is equally negligible. Both mechanisms can therefore be excluded, and the measured damping is governed instead by defect- and surface-related dissipation, which we both model within the framework of two-level systems (TLS)~\cite{phillipsTwolevelStatesGlasses1987}.

Microscopic defects in materials are commonly modeled as asymmetric double-well potentials, characterized by an asymmetry energy $\Delta$ and a barrier height $V_0$ separating the two wells~\cite{phillipsTwolevelStatesGlasses1987}.
TLS have traditionally been studied in the quantum-tunneling regime, where relaxation proceeds via well-to-well tunneling and which dominates only at sub-kelvin temperatures~\cite{phillipsTwolevelStatesGlasses1987}. In our experiments, spanning \SIrange{4}{300}{\kelvin}, the relevant dynamics is instead thermally activated over the barrier $V_0$.

Following the approach of Ref.~\cite{hamoumiMicroscopicNanomechanicalDissipation2018}, we describe the aggregate mechanical response using two distinct sets of distributions of $\Delta$ and $V_0$. The first describes a single, well-defined defect configuration~\cite{hunklinger3UltrasonicProperties1976}, which we use to model losses originating within the crystalline \ce{MoS2} lattice itself; the crystallinity of \ce{MoS2} justifies this single-defect picture even for a monolayer. The second employs a broad distribution of $\Delta$ and $V_0$ originally established for amorphous solids~\cite{vacherAnharmonicRelaxationalSound2005}, which we adopt to describe the structurally disordered environment at the \ce{MoS2} surface.

For the same monolayer drum as in Fig.~\ref{fig:fig5}a, we plot the inverse intrinsic quality factor $Q^{-1}_\mathrm{int}$ as a function of temperature in Fig.~\ref{fig:fig5}b. We fit the combined TLS model introduced above to this data, using
$Q^{-1}_\mathrm{int} = Q^{-1}_\mathrm{TLS,single} + Q^{-1}_\mathrm{TLS,dist}$,
where $Q^{-1}_\mathrm{TLS,single}$ and $Q^{-1}_\mathrm{TLS,dist}$ denote the contributions of the single defect and the defect distribution in the thermally activated regime, respectively. As the fit carries a large number of free parameters, we fix the attempt time $\tau_0$ to the inverse Debye frequency, $\tau_0 = \SI{0.183}{\pico\second}$~\cite{pengThermalConductivityMonolayer2016}, consistent with the common assumption that the attempt frequency is of order the Debye frequency~\cite{phillipsTwolevelStatesGlasses1987}. The resulting fit is shown as a green solid line in Fig.~\ref{fig:fig5}b, with the single-defect and defect-distribution contributions shown as blue and turquoise dashed lines, respectively. It reproduces the measured damping rate across the full temperature range. The fit values we obtain carry considerable uncertainty and serve rather as an indication; a more statistically robust analysis would require either a better microscopic understanding of the underlying defects or a more extensive dataset, both of which lie beyond the scope of this work. Details of the fitting procedure are provided in Supplementary Section~\ref{SI:Damping}.
 
Next, we contrast the monolayer results with data obtained on a multilayer drum with a thickness of \SI{24}{\nano\meter} ($37$ layers) and a radius of \SI{5}{\micro\meter}. The intrinsic quality factor is obtained by first subtracting the extrinsic clamping loss from the measured quality factor to obtain the diluted intrinsic quality factor (see Fig.~\ref{fig:fig4}b) before applying the dissipation dilution model using the procedure described in Fig.~\ref{fig:fig5}a. Figure~\ref{fig:fig5}c shows the inverse intrinsic quality factor as a function of temperature. Measurements are limited to temperatures up to \SI{50}{\kelvin}, as the lower quality factor of the multilayer drum precludes sufficiently accurate measurements at higher temperatures.
We repeat the analysis of the underlying intrinsic damping mechanisms described above. The resulting fit for the multilayer is included in Fig.~\ref{fig:fig5}c as a green solid line, with the single-defect and defect-distribution contributions shown as blue and turquoise dashed lines, respectively. As for the monolayer, the fit closely reproduces the measured damping rate, though it carries the same caveats regarding parameter uncertainty discussed above (Supplementary Section \ref{SI:Damping}).

For both the monolayer and multilayer drum, the distributed contribution (turquoise) dominates the damping at low temperature, while the single-defect contribution (blue) takes over only at higher temperature, in agreement with Ref.~\cite{hamoumiMicroscopicNanomechanicalDissipation2018}. This crossover reflects the thermally activated nature of the single-defect relaxation: governed by one fixed barrier $V_0$, its relaxation time depends exponentially on $V_0/T$ and becomes far longer than the mechanical period at low temperature, effectively freezing out this channel. The distributed contribution, spanning a wide range of barrier heights, always retains a subpopulation of low-barrier defects able to relax on the timescale of the oscillation, sustaining a finite contribution even at cryogenic temperatures.

Consistent with Fig.~\ref{fig:fig4}, $Q_\mathrm{int}^{-1}$ is systematically higher for the monolayer than for the multilayer drum.
This trend, together with the relative weight of the two damping channels, reflects the far larger surface-to-volume ratio of the monolayer: its damping is dominated by the distributed (surface) contribution, while the thicker multilayer is dominated by the single-defect (bulk) contribution.
The distributed contribution extends to barrier heights of $V_1$ $\approx$ \SI{1.3}{eV} (\SI{0.04}{eV}) for the monolayer (multilayer).
The monolayer values are compatible with the eV-scale energies of adatoms,
vacancies, and adsorbates \cite{zhaoStabilityDefectsMonolayer2017,amaniNearunityPhotoluminescenceQuantum2015}, consistent with the distributed
contribution capturing such surface-related defects. For the single-
defect contribution, we obtain barrier heights of $V_0 \leq$ \SI{67}{meV} for both drums. This is too small to be consistent with the mentioned defect hopping. Tentatively, a more plausible candidate at this energy scale is a defect configuration related to the Jahn-Teller effect, known to couple to strain \cite{jansenTipinducedCreationJahnTeller2024}, which is beyond the scope of the current work.

\subsection*{Conclusions and outlook}
In summary, we report a new benchmark for quality factors in van der Waals mechanical resonators, reaching a quality factor close to $400{,}000$ in a monolayer \ce{MoS2} drum at \SI{4}{\kelvin}, nearly an order of magnitude beyond the previous best reported for any 2D material in this temperature range.

The monolayer drums studied here operate at a tension of $\SI{0.22}{\newton/\meter}$, which translates to a tensile stress of $ \SI{0.34}{\giga\pascal}$, roughly a third of the $\sim\SI{1}{\giga\pascal}$ employed in comparable strained-silicon nitride or silicon nanostrings~\cite{villanuevaEvidenceSurfaceLoss2014,beccariStrainedCrystallineNanomechanical2022}. This leaves substantial room for improvement: the reported breaking strength of \ce{MoS2} of up to \SI{15}{\newton\per\meter}~\cite{bertolazziStretchingBreakingUltrathin2011} corresponds to a stress of $\sim\SI{23}{\giga\pascal}$ at monolayer thickness, roughly an order of magnitude above typical fracture strengths of amorphous \ce{Si3N4} thin films and more than a factor of $20$ above the operating stress of state-of-the-art top-down devices. This reflects the near-ideal, defect-sparse nature of atomically thin crystals, which can approach their theoretical strength limit rather than the flaw-limited strength of macroscopic films. Larger tensile stress in our devices could be achieved through electrostatic gating, substrate engineering for enhanced thermal differential straining, or strain engineering of the drum geometry itself; we estimate that dilution factors of order $8,000$ are achievable for a \SI{5.5}{\micro\meter} radius monolayer drum.
Such quality factors would benefit a range of applications, including quantum transducers approaching unity efficiency~\cite{gaoHighspeedQuantumTransducer2020}, magnetoresistive sensors in related van der Waals materials such as \ce{WTe2}~\cite{leeSingleFewlayerWTe22016}, and THz photonics~\cite{zalalutdinovAcousticCavities2D2021}.

Looking ahead, the extreme dissipation dilution demonstrated here suggests that the concepts driving the ongoing revolution in top-down nanomechanical Q, soft clamping~\cite{tsaturyanUltracoherentNanomechanicalResonators2017} and strain engineering~\cite{ghadimiElasticStrainEngineering2018}, could translate powerfully to van der Waals materials. The thinnest top-down platform demonstrated to date, single-crystal strained silicon at \SI{14}{\nano\meter}, already combines dissipation dilution with soft clamping to reach quality factors exceeding $10^{10}$~\cite{beccariStrainedCrystallineNanomechanical2022}; applying comparable strategies to multilayer \ce{MoS2}, at similar or smaller thickness, could push dilution factors substantially beyond those reported here. 
A further strategy to enhance the dissipation-diluted quality factor is to cool beyond the temperature range studied in this work. At \SI{}{\milli\kelvin} temperatures both a further increase of the thermally induced tension and a decrease of defect-mediated dissipation might favorably affect the quality factor \cite{willHighQualityFactor2017,guttingerEnergydependentPathDissipation2017}.

Beyond these, van der Waals materials offer a further advantage largely absent from established top-down platforms: atomic-scale defects whose position can be deterministically engineered. Combined with the ultrahigh $Q$ demonstrated here, this positions \ce{MoS2} as a promising platform for spin-mechanical hybrid systems, in which engineered atomic defects could be coherently coupled to a mechanical mode of exceptional coherence.

\subsection*{Methods}
\textbf{Sample fabrication}
We fabricate suspended \ce{MoS2} drums by transferring exfoliated mono- and few-layer flakes onto pre-structured Si/SiO$_2$ substrates (\SI{525}{\micro\meter}/\SI{285}{\nano\meter}). Circular cavities with radii between \SI{2.5}{\micro\meter} and \SI{6}{\micro\meter} and a depth of approximately \SI{1500}{nm} are defined by optical lithography and anisotropic dry etching using a \ce{CHF3} and \ce{SF6} gas mixture. The substrate is subsequently coated with a thin layer of Ti/Au (\SI{2}{\nano\meter}/\SI{5}{\nano\meter}); while the Ti serves as an adhesion layer, the Au is required for the transfer of the flakes. The \ce{MoS2} flakes are transferred onto the metalized surface using the gold-assisted direct exfoliation technique~\cite{velickyMechanismGoldAssistedExfoliation2018, liuDisassembling2DVan2020, geilenSituMappingPressureInduced2025}, followed by annealing at \SI{200}{\degreeCelsius} in ambient conditions. This transfer technique yields membranes with exceptionally low residue contamination.

\textbf{Photoluminescence imaging}
The Photoluminescence (PL) image overlay is created by capturing a photograph using a $10$x Mitutoyo M Plan ($0.28$ NA) objective without any filters, followed by a second image using a \SI{550}{nm} short pass filter in front of the illumination LED and \SI{600}{nm} long pass filter in front of the camera. The PL is further enhanced by replacing the beamsplitter with a \SI{550}{nm} dichroic mirror. The near-IR filter in front of the CMOS camera chip has also been removed.

\textbf{Mechanical characterization}
Measurements are performed in an attoDRY800 cryostat with a chamber pressure below \SI{e-4}{\pascal}. Prior to measurements, the sample is kept under vacuum for several days to allow pressure equilibration across the drum. Without this precaution, the pressure differential between the sealed volume beneath the drum and the rest of the vacuum chamber can induce blistering of the flake~\cite{geilenSituMappingPressureInduced2025} - an effect which can last for several days. Mechanical motion is actuated by a piezoelectric shaker and read out optically using a Koheras BasiK E15 fiber laser at \SI{1550}{\nano\meter}. The reflected light, modulated by the drum's out-of-plane displacement, is detected by a photodetector and demodulated using a Zurich Instruments HF2LI lock-in amplifier. The mechanical response is obtained by sweeping the piezo drive frequency through resonance. Quality factors are extracted independently via ringdown measurements, in which the drive is switched off at resonance, and the subsequent amplitude decay is recorded to yield the energy decay rate $\Gamma = 2\pi f_0 / Q$.

Spatial mapping of the mechanical modes is enabled by a micropositioning system that scans the sample with respect to the laser spot, which is focused to below \SI{1}{\micro\meter} (diffraction-limited at \SI{1550}{\nano\meter}). This allows individual drums across the chip to be addressed and their mode shapes to be imaged, confirming the origin of each resonance signal.

\textbf{Determination of the intrinsic quality factor}
The intrinsic quality factor of monolayer drums at cryogenic temperature is obtained via two distinct methods:
Figure~\ref{fig:fig4}a presents an ensemble analysis across multiple drums of varying radius at \SI{4}{\kelvin}, yielding an average $Q_\mathrm{int} = 212 \pm 10$. 
The temperature-dependent analysis in Fig.~\ref{fig:fig5}a includes data for a single-monolayer drum. Even more, Fig.~\ref{fig:fig5}a comprises data between room temperature and \SI{60}{\kelvin}. Lower temperatures are not included in the analysis of Fig.~\ref{fig:fig5}a, as the tension starts to saturate at this temperature, while the $Q$-factor keeps increasing, as apparent in Fig.~\ref{fig:fig3}a and b, respectively. Incorporating this data in Fig.~\ref{fig:fig5}a would thus produce a vertical smearing of the data points and obscure the analysis. 
The modest difference between these value of $Q_\mathrm{int} = 186 \pm 19$ (Fig.~\ref{fig:fig5}a) and $Q_\mathrm{int} = 212 \pm 10$ (Fig.~\ref{fig:fig4}a) is not only attributed to drum-to-drum variability but also reflects the change of the damping rate between \SI{60}{\kelvin} and base temperature (see also Fig.~\ref{fig:fig3}c).

\newpage

\printbibliography[heading = subbibliography]
\end{refsection}

\textbf{Data availability:} The data that support the findings of this study are available at Zenodo XXX. 

\subsection*{Acknowledgements}
We gratefully acknowledge financial support from the Deutsche Forschungsgemeinschaft (DFG, German Research Foundation) under Germany’s Excellence Strategy, EXC-2111-390814868  and Munich Quantum Valley K6.

\subsection*{Author contributions}
L.G., A.M. developed the transfer process and L.S., L.G., A.M. fabricated the samples. The samples were characterized by L.S. and A.R. with assistance of P.B. Data were analyzed by L.S. and interpreted by L.S. and E.M.W. with assistance from other authors. Theory was developed by L.S. and E.M.W. The manuscript was written by all authors.
The work was supervised by A.W.H. and E.M.W.

\subsection*{Competing interests}
The authors declare no competing interests.  

\subsection*{Additional information}
\textbf{Supplementary information} The online version contains supplementary material available at 

\textbf{Correspondence and requests} for materials should be addressed to E.M. Weig.

\onecolumn
\appendix

\end{document}